\documentclass[letterpaper]{article}
\usepackage{verbatim,enumerate,balance}
\usepackage{graphicx}
\usepackage{amssymb}
\usepackage{amsmath}
\usepackage{amsfonts}
\usepackage{pseudocode,algorithm,algorithmicx,newalg}
\begin{document}
\title{Two Way Concurrent Buffer System without Deadlock in Various Time Models Using Timed Automata}
\author{Rohit Mishra, Md Zeeshan and Sanjay~Singh\thanks {Sanjay Singh is with the Department of Information and Communication Technology, Manipal Institute of Technology, Manipal University, Manipal-576104, INDIA \hspace{5cm}E-mail: sanjay.singh@manipal.edu}}

\maketitle

\begin{abstract}
Two way buffer system is a system that exhibits transfer of data using two buffers concurrently. It includes processes that synchronize to exchange data with each other along with executing certain delays between these synchronizations. In existing Tiny Two Way Buffer System, both generators produce packets in half duplex manner in no time, deterministic time, and non deterministic time. Analysis of the model for above time options leads the model in deadlock. The model can be out of the deadlock if timings in the model is incorporated in alternative fashion. The generators produce packets after a delay of 10 seconds. However, generator one has an initial shift of 5 seconds after which it begins sending a packet every 10 seconds. Hence, initial delay for generator one is 15 seconds and for generator two it is 10 seconds. Due to this initial shift, both generators produce packets alternatively and is deadlock free as the packets do not meet at the same time instant. Moreover, the existing system model is not concurrent and hence takes more time for packet transfer in every iteration. In this paper we have proposed a model of buffer system using an additional dummy buffer for transfer of data packets, we thus checking the model with various time models as no time, deterministic time and non deterministic time. The results of proposed model under above time models are in deadlock. We achieve deadlock free situation by introducing appropriate delay in various buffers of the proposed system, the delay timing is nondeterministic time. The new approach speeds up the transfer of packets, as a result the transfer of data becomes concurrent, deadlock free and hence the model proposed is time efficient. To model and simulate the proposed system we have used UPPAAL as a model checking tool environment for modeling, validation and verification of real-time systems modeled as networks of timed automata. Simulation results shows that the proposed two way buffer system is fully concurrent and time efficient as compared to the existing buffer system.
\end{abstract}


\section{Introduction}
Industrial embedded systems are hardware systems that have associated software to operate them and which perform specific functionality. Such systems are often highly concurrent with several parallel components operating together \cite{AS11}. These components have synchronizations and dependencies between them that make such systems highly complex. The formal approaches are used to design models of completely new systems as well as to evaluate the existing systems. The correctness of these models becomes very crucial since they form input for the design of the actual software that controls the operations of the system hardware. Simulation is one approach that allows the analysis of certain behavioral aspects of the system. However, it is also essential to ensure that the system behaves accurately under all possible circumstances. 
\par Embedded systems with hard real time constraints are very time-specific and require the verification of time-specific logistic rules. In such cases, model checking approaches that also allow time based verification needs to be considered. The main purpose of this work is to perform time based model checking on such an industrial system. 
	
\par However the accurate time-based verification requires the translation of the current design models of the system to formal semantics of existing modeling languages and analysis of the entire state space of the system by means of state-of-the-art model checkers. Several model checkers are suitable for this kind of verification such as UPPAAL \cite{GDK04}\cite{JKF95}, mCRL2 \cite{JAMY06} and other temporal logic model checkers \cite{WIKI12}. These model checkers can be used to model highly concurrent systems and perform verification of required properties. In this paper we have used UPPAAL as  model checking tool which allows  verification of the required properties on the system for all its dynamic behaviors. This paper is the extended version of our earlier work \cite{REF13}.
\par
The remaining paper is organized as follows. Section \ref{1} explains the concept of timed automata. Section \ref{2} briefly describe about the various modeling approaches. Section \ref{3} explains various time models which affects the deadlock condition. Section \ref{4} describes the model checking tool UPPAAL. Section \ref{5} examines the limitation of the existing tiny two way buffer system. Section \ref{6} discusses about the proposed two way buffer system.  Section \ref{7} explains about the time analysis of proposed system. Section \ref{8} describes about the UPPAAL model of proposed buffer system in no time. Section \ref{9} describes about the UPPAAL model of proposed buffer system in deterministic time. Section \ref{10} describes about the UPPAAL model of proposed buffer system in non-deterministic time for no deadlock. Section \ref{11} discusses about buffer sizing of the proposed system which affects the efficiency of the overall system. Section \ref{12} describes the scope of the work and the future enhancement. Finally, section \ref{13} concludes this paper.

\section{Timed Automata}
\label{1}
A timed automaton (TA) is a finite-state machine extended with clock variables. It uses a dense-time model where a clock variable evaluates to a real number \cite{PUK12}. All the clocks progress synchronously. A timed automaton manipulate clocks, evolving continuously and synchronously with absolute time.\\
 The elements of TA include:
    
\begin{itemize}
	\item Each transition of such an automaton is labeled by a constraint over clock values (also called guard), which indicates when the transition can be fired
	\item A set of clocks to be reset when the transition is fired
	\item Each location is constrained by an invariant, which restricts the possible values of the clocks for being in the state, which can then enforce transition to be taken
	\item The time domain can be $\mathbb{N}$, the set of non-negative integers, or $\mathbb{Q}$, the set of non-negative rationals, or even $\mathbb{R}$, the set of non-negative real numbers
\end{itemize}
 
\textbf{Definition: Timed Automaton}: A timed automaton is a tuple (L, $l_0$, C, A, E, I) where,
\begin{itemize}
	\item L is a set of locations,
	\item $l_0 \in L$ is the initial location,
	\item $C$ is the set of clocks,
	\item $A$ is a set of actions, co-actions and the internal $\tau$-action,
	\item $E \subseteq L \times A \times B(C) \times 2^{C} \times L$ is a set of edges between locations with an action,
	\item $B(C)$ is a set of conjunctions over simple conditions of the form $x \bowtie c$ or $x - y \bowtie c$, where $x,y \in C$, $c \in \mathbb{N}$ and $\bowtie \in \{<, \geq, =, \leq, >\}$
	\item a guard and a set of clocks to be reset and
	\item $I : L \rightarrow B(C) $ assigns invariants locations.
\end{itemize}

\section{Various Modeling approaches}
\label{2}
These days there are many model checking tools used to model and adopt different aspects towards verification. SMV is the first model checker to use Binary Decision Diagrams (BDDs) \cite{REF10}, which are data structures used to represent boolean functions. mCRL2 \cite{REF8} is a model checker based on a specification language for describing concurrent discrete event systems and which uses mu-calculus \cite{REF9} for its specifications. Spin is used to perform verification on distributed systems and uses Linear Temporal Logic (LTL) to specify its properties \cite{REF11}. HyTech is designed for reasoning about temporal requirements in hybrid systems \cite{REF12}.
 \par Temporal logic model checkers are most reliable to perform model checking on the machines where the properties to be verified are time specific. The properties stating that an event can eventually happen or always happen after a given period of time is possible through temporal logic. Using temporal logic it is possible to represent time specific properties crucial for real time system we have designed. 
\par There are various model checking tools that are used to perform the kind of verification for the two way buffer system which is a  real time system. Uppaal is the model checker for real time systems that uses the concept of timed automata for modeling purposes and a variant of Computation Tree Logic (CTL) for its specifications \cite{REF7}.

\section{Various Time Models}
\label{3}
The verification of the system requires the evaluation of various multiple time models as follows:
\begin{itemize}
	\item No time: It is a model of the system that does not consider the time taken by different actions and mainly considers the various possible sequences of interactions between the components. Verification on a system with no time ensures that all possible scenarios are considered and hence a system which satisfies a property with no time always satisfies the property.
	\item Deterministic time: In this model, specific time durations are assigned to all actions and the system is analyzed for this timing behavior. With deterministic time only a particular time sequence or behavior is considered. If a property is violated in the untimed version of the system, the system is not necessarily incorrect. This is because the property might be satisfied in the deterministic model of the system for all realistic time sequences which can occur in the actual machine.
	\item Non-deterministic time: This model allows actions to occur non deterministically within time intervals. This allows verification to be performed on a system with variable timings within given realistic ranges. Since the industrial systems have timing involved in them and the time duration of activities can vary, the non-deterministic models with time give the most realistic models of the industrial
machines. However, since the timing variations are within very small limits, the deterministic models can be efficiently used to analyze the behavior of the system.
\end{itemize}
        
\section{UPPAAL - Model Checking Tool}
\label{4}
In TA, notion of time is introduced by clock variables, which are used in clock constraints to model time-dependent behavior. Systems comprising multiple concurrent processes are modeled by networks of timed automata, which are executed with interleaving semantics and synchronization on channels. UPPAAL  is a tool set for the modeling, simulation, animation and verification of networks of timed automata. The UPPAAL model checker enables the verification of temporal properties, including safety, liveness and reachability properties. The simulator can be used to visualize counterexamples produced by the model checker in case the given specification is not satisfied by the model.
\par The UPPAAL modeling language extends timed automata by introducing bounded integer variables, binary and broadcast channels, and urgent and committed location. Timed automata are modeled as a set of locations, connected by edges. The initial location is denoted by $\circledcirc$. Invariants can be assigned to locations which enforce synchronization between them. Edges may be labeled with \textit{guards}, \textit{synchronizations} and \textit{updates}. Updates are used to reset clocks and to manipulate the data space. Processes synchronize by sending and receiving events through channels. Sending and receiving via a channel 'c' is denoted by c! and c?, respectively. Binary channels are used to synchronize one sender with a single receiver. A synchronization pair is chosen non-deterministically if more than one channel is enabled. Broadcast channels are used to synchronize one sender with an arbitrary number of receivers. Any receiver that can synchronize must do so. Urgent and committed locations are used to model locations where no time may pass. Committed locations enforce that the synchronization is atomic and are denoted by the symbol $\copyright$. Leaving a committed location has priority over leaving non-committed locations.
\par A UPPAAL model comprises three parts: \textit{global declarations}, \textit{parameterized timed automata} and a \textit{system declaration}. In the global declarations section, global variables, constants, channels and clocks are declared. In the system declaration, TA templates are instantiated and the system to be composed is given as a list of TA.

\section{Existing Tiny Two Way Buffer System}
\label{5}
The existing system consists of two generators, two buffers, and two exits as shown in Figure 1 \cite{AS11}. The first generator G1 repeatedly produces packets. This packet is received over channel 'a' by buffer M1 which sends it over channel 'c2'. The packet is received over this channel by a second buffer M2 which again sends it over channel 'e' to be received by the exit E2. Similarly, the second generator G2 repeatedly generates packets and sends them over channel 'b'. This is received by buffer M2 which sends it over channel 'c1' to buffer M1. M1 again forwards the packet over channel 'd' to exit E1. This system, obviously, has a deadlock which occurs when both G1 and G2 produce packets at the same instant of time. 
\par In that case, the packet is received by the buffers M1 and M2 respectively at the same time. As a result, both M1 and M2 wait to send packets to each other and the system deadlocks.
\par Verification of the system has been done for various time models as for no time, deterministic time and non deterministic time. and it is giving significance results as for no time, deterministic time and nondeterministic time deadlock property not satisfied in the Uppaal user interface ie deadlock occurs with no time, deterministic time and nondeterministic time.
\par In order to study a system with no deadlock, a slightly modified version of the tiny two way buffer system was considered \cite{AS11}. In this system, both generator G1 and G2 produce a packet after a delay of 10 seconds. However, G1 has an initial shift of 5 seconds after which it begins sending a packet every 10 seconds.
Hence, initial delay for G1 is 15 seconds and for G2 it is 10 seconds. Due to this initial shift of 15 seconds at G1 and 10 seconds delay at G2 the time consumption in waiting to maintain synchronization in transfer of packets is more.
\begin{figure}[bpht!]
\centering
	\includegraphics[scale=0.25]{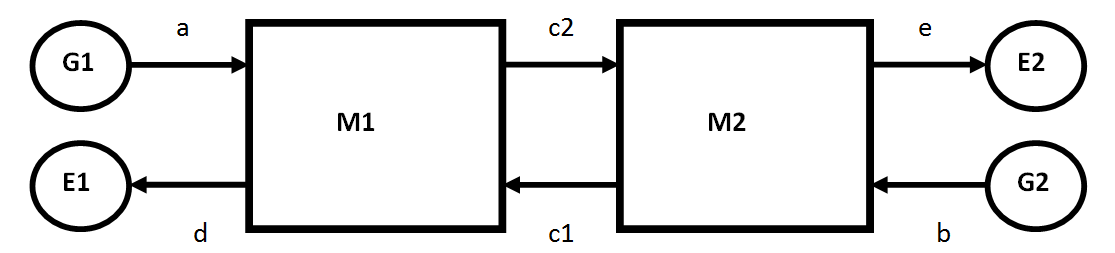}
	\caption{Existing Model of Two Way Buffer System}
	\label{Fig1}
\end{figure}

\section{Proposed Two Way Buffer System}
\label{6}
In this paper we have proposed a two way buffer system which consists of two generators, three buffers and two exits. The generators generate the packets to be transferred until it reaches the exits. Figure 2 depicts the proposed model. 

\begin{figure}[bpht!]
\centering
	\includegraphics[scale=1.4]{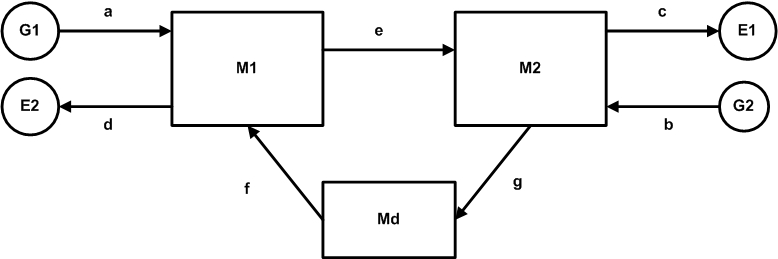}
	\caption{Proposed Model of Two Way Buffer System}
	\label{Fig2}
\end{figure}
	
First, the generator G1 repeatedly produces packets. This packet is received over the channel 'a' by buffer M1. The packet is received by buffer M2 over channel 'e' and sends it to exit E1 via the channel 'c'. Similarly, the second generator G2 repeatedly generates packets and sends them over channel 'b'. This is received by buffer M2 which sends it over an intermediate dummy buffer Md through channel 'g'. The dummy buffer sends the packet to buffer M1 via channel 'f'. M1 again forwards the packet over channel 'd' to exit E1. This process is repeated infinitely until it is stopped. We are incorporating the delay $(\theta)$ in transmission of packet at buffer M1 and waiting for M2 to be free. After waiting, buffer M1 will release the packet and send it to M2. On the other direction the packet released from M2 will reach at buffer Md, now Md will wait until M1 is free. Then Md will release the packet and send it M1. Based on the clocks being synchronized the data is transferred concurrently. The two way data packet transfer through the proposed system is concurrent. The deadlock is avoided as there is a continuous synchronization of time between each of the entities modeled. \\ 
The Timed Automata representation of the proposed two way concurrent buffer system is shown in Fig 3. 

\begin{figure}[bpht!]
\centering
	\includegraphics[scale=1.4]{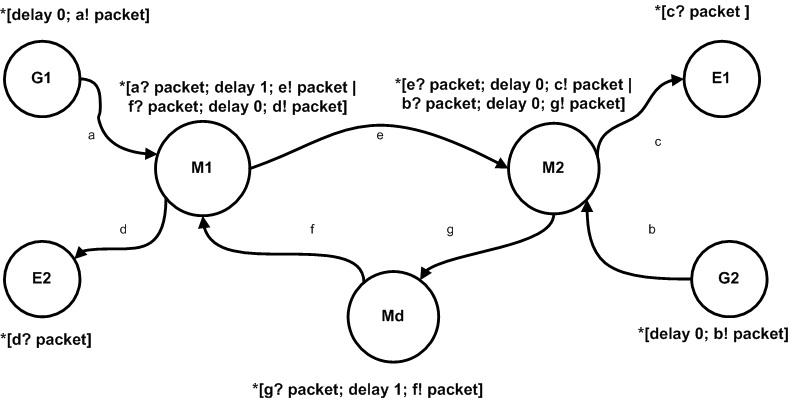}
	\caption{Timed Automata Representation of Proposed Two Way Buffer System}
\end{figure}
	
%



\section{Timing Analysis}
\label{7}
In this section, the timing analysis of individual buffers, entry and exit points has been done.\\
Notations used:
\begin{itemize}
	\item $\zeta$ : Time required over the channel for transfer of packets from one location to next consecutive location
	\item $\theta$ : Delay in transmission time of packet at buffer M1 and Md
	\item $\alpha$ : Iteration time for every transfer of packet from generator to exit in both the directions
\end{itemize}

At G1, time required to move the packet from G1 to M1 is $\zeta$ time units. At M1 it will wait for $\theta$ time units, because it has to wait for M2 to be free, after waiting $\theta$ time units M2 will be  free. So, total time at M1 is $\zeta+\theta$ time units. Time to reach from M1 to M2 will take $\zeta$ time units, so the total time at M2 is 2$\zeta+\theta$ time units.  The time to reach from M2 to E1 is $\zeta$ time units, so the total time at exit point E1 is 3$\zeta+\theta$ time units. Thus the time taken for transfer of packet from G1 to E1 is 3$\zeta+\theta$ time units. Additional delay of $\zeta$ time units is added to perform synchronization with G2. For next iteration G1 will release the packet at 4$\zeta+\theta$ time units.
\par On the other end at entry point G2 the packet is ready at $0^{th}$ time unit. Time taken by a packet to move from G2 to M2 requires $\zeta$ time units, total time required at M2 is $\zeta$ time unit. From M2 to Md , the time required is $\zeta$ time units, and there is a delay in the buffer Md, so total time at Md is 2$\zeta+\theta$ time units. At Md it will wait for $\theta$ time units. Because it has to wait for M1 to be free, after waiting $\theta$ time units M1 will be  free. The time taken to move the packet from Md to M1 is $\zeta$ time unit, now total time at M1 is 3$\zeta+\theta$. On the other side M1 will move the packet to exit E2 and the time taken is $\zeta$ time unit, now the total time at E2 is 4$\zeta+\theta$ time units. At this point on completion of one cycle G2 is ready at 4$\zeta+\theta$ time units. In this way the synchronization is maintained because G1 and G2, releasing the packets at 4$\zeta+\theta$ time units in either direction. Also the concurrency is maintained as packet transfer is concurrent in both the directions throughout the system.\\
Below is the mathematical form of timing analysis and certain assumptions:\\
Assumptions: Let $\alpha$ be a iteration number, $\theta$ be the delay at buffer and $\zeta$ be the delay duration of channel. 
 
\begin{itemize}
	\item Time for subsequent packet transfer at generators G1 or G2 is:
\\ For $\alpha$ = 0, the packet is ready at 0 time units;
\\ For $\alpha$ = 1, time consumed is 4$\zeta+\theta$ time units;
\\ For $\alpha$ = 2, time consumed is 8$\zeta+2\theta$ time units;
\\ In general $\displaystyle\sum\limits_{\alpha=0}^{\infty} [\alpha(4\zeta+\theta)]$.\\

At G2, first packet will transfer without any delay at $3\zeta+\theta$, but one additional delay is provided to synchronize with G2. 

\item When M1 receives a packet from G1, the timing analysis is done as follows:
\\ For $\alpha$ = 0, time consumed is $\zeta+\theta$ ;
\\For $\alpha$ = 1, time consumed is $5\zeta+2\theta$ units;
\\For $\alpha$ = 2, time consumed is $9\zeta+3\theta$ units;
\\ In general $\displaystyle\sum\limits_{\alpha=0}^{\infty} [\zeta+\theta+\alpha(4\zeta+\theta)]$ 

\item When M1 receives a packet from Md, the timing analysis is done as follows:
\\ For $\alpha$ = 0, time consumed is $3\zeta+\theta$ ;
\\For $\alpha$ = 1, time consumed is $7\zeta+2\theta$ units;
\\For $\alpha$ = 2, time consumed is $11\zeta+3\theta$ units;
\\ In general $\displaystyle\sum\limits_{\alpha=0}^{\infty} [3\zeta+\theta+\alpha(4\zeta+\theta)]$ 

\item When M2 receives a packet from M1, the timing analysis is done as follows:
\\ For $\alpha$ = 0, time consumed is $2\zeta+\theta$ ;
\\For $\alpha$ = 1, time consumed is $6\zeta+2\theta$ units;
\\For $\alpha$ = 2, time consumed is $10\zeta+3\theta$ units;
\\ In general $\displaystyle\sum\limits_{\alpha=0}^{\infty} [2\zeta+\theta+\alpha(4\zeta+\theta)]$ 

\item When M2 receives a packet from G2, the timing analysis is done as follows:
\\ For $\alpha$ = 0, time consumed is $\zeta$ ;
\\For $\alpha$ = 1, time consumed is $5\zeta+\theta$ units;
\\For $\alpha$ = 2, time consumed is $9\zeta+2\theta$ units;
\\ In general $\displaystyle\sum\limits_{\alpha=0}^{\infty} [\zeta+\theta+\alpha(4\zeta+\theta)]$ 

\item When Md receives a packet from M2, the timing analysis is done as follows:
\\ For $\alpha$ = 0, time consumed is $2\zeta+\theta$ ;
\\For $\alpha$ = 1, time consumed is $6\zeta+2\theta$ units;
\\For $\alpha$ = 2, time consumed is $10\zeta+3\theta$ units;
\\ In general $\displaystyle\sum\limits_{\alpha=0}^{\infty} [2\zeta+\theta+\alpha(4\zeta+\theta)]$ 

\item When E1 receives a packet from M2, the timing analysis is done as follows:
\\ For $\alpha$ = 0, time consumed is $3\zeta+\theta$ ;
\\For $\alpha$ = 1, time consumed is $7\zeta+2\theta$ units;
\\For $\alpha$ = 2, time consumed is $11\zeta+3\theta$ units;
\\ In general $\displaystyle\sum\limits_{\alpha=0}^{\infty} [3\zeta+\theta+\alpha(4\zeta+\theta)]$ 
\item When E2 receives a packet from M1, the timing analysis is done as follows:
\\ For $\alpha$ = 0, time consumed is $4\zeta+\theta$ ;
\\For $\alpha$ = 1, time consumed is $8\zeta+2\theta$ units;
\\For $\alpha$ = 2, time consumed is $12\zeta+3\theta$ units;
\\ In general $\displaystyle\sum\limits_{\alpha=0}^{\infty} [4\zeta+\theta+\alpha(4\zeta+\theta)]$
\end{itemize}

The pseudocode for the system proposed:

\begin{verbatim} 
// Initial state is current state
while(new_packet=TRUE)
  check(delay) 
    if(buffer.next = TRUE)
      delay = 0  //Assigning zero delay
      if(delay = = 0)
        transition(l, l^i)
        l=l'
        l'=next_location
        if(buffer.next = free)
          decrement_delay
          check(delay) 
          
\end{verbatim}

Here $l$, $l^{i}$ are the locations as per the timed automata formalism. $\textit{Check()}$ function verifies if there is any packet on the opposite way being transferred then it will delay the packet in the corresponding buffer.

\section{Uppaal Model of Proposed  Buffer System without Time}
\label{8}
This kind of model of the system has no delay incorporated as there is no time. Hence, the delay variables are all null in this system. This system has a deadlock when both generators produce a packet at the same time which are received by M1, M2 and Md. The buffers wait for each other to be free so they can forward their packets and the system deadlocks.  The location marked with red indicates where the control is at the moment.

\subsection{} Generators G1 and G2 has no delay so they do not wait. as the new packet generates it transfers that to next channel. as shown in the Fig 4.
\begin{figure}[bpht!]
\centering
	\includegraphics[scale=0.75]{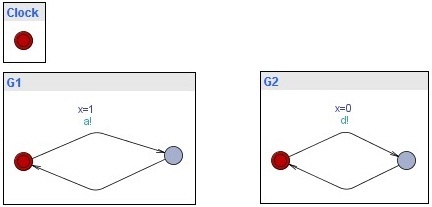}
		\caption{Generators G1 nad G2 in no time}
	\label{Fig 4}
\end{figure}

\subsection{} Similar to generators, buffers will transfer the packets to next channel as they received it without any delay. as shown in the Fig 5.
\begin{figure}[bpht!]
\centering
	\includegraphics[scale=0.75]{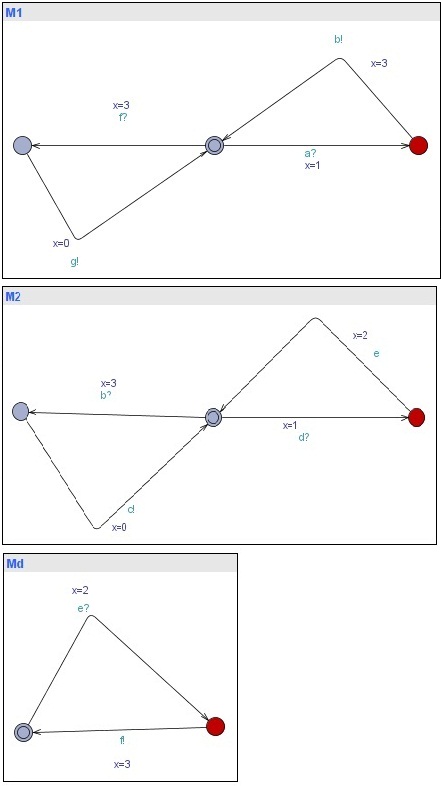}
		\caption{Buffers M1, M2 and Md in no time}
	\label{Fig5}
\end{figure}

\subsection{} Similar to Generators and buffers Exits also forward the packets as they receive it. as shown in the Fig 6.
\begin{figure}[bpht!]
\centering
	\includegraphics[scale=0.75]{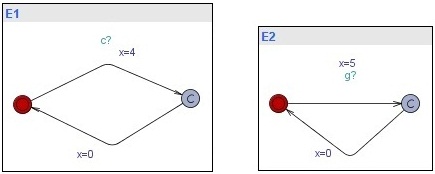}
		\caption{Exit E1 and E2 in no time}
	\label{Fig 6}
\end{figure}

\subsection{} Verifier gives the result for deadlock, because "no time" approach violates the Proposed two way concurrent buffer system timings where we require slight delay at various positions to make the system concurrent. as shown in the Fig 7.
\begin{figure}[bpht!]
\centering
	\includegraphics[scale=0.70]{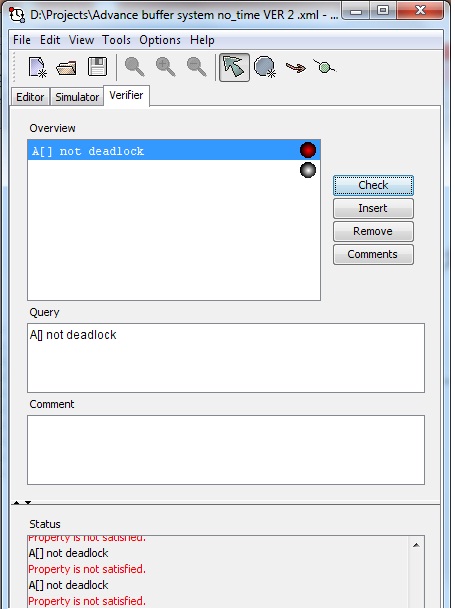}
		\caption{verification result for no time}
	\label{Fig 7}
\end{figure}

\section{Uppaal Model of Proposed Buffer System in Deterministic Time}
\label{9}
In this section, some deterministic time is added to the system. The generator, G1, produces a packet every $\alpha$ seconds and the generator G2 produces packets every $\beta$ seconds. Also, after receiving a packet the intermediate buffers M1,M2 and Md delay for $\gamma$ seconds each before forwarding the packets to the respective exits. This process repeats over time. One possibility is that $\alpha$ is delay 10, $\beta$ is delay 1 and $\gamma$ is delay 2. It can be observed easily that there exists a deadlock in the system at time 10 seconds when both generators produce a packet at the same time.

\subsection{} Generators G1 and G2 will transfer the packets only when clock hits particular time not before and not after. so giving delay with this time system is logically difficult. in Fig 8 below transfer of packets will occur only at 0th and 5th second.
\begin{figure}[bpht!]
\centering
	\includegraphics[scale=0.75]{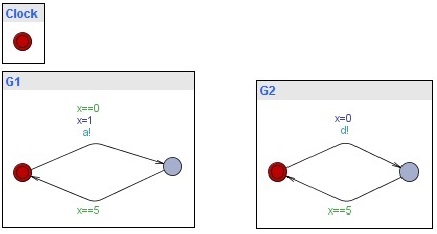}
		\caption{Generators G1 and G2 in deterministic time}
	\label{Fig 8}
\end{figure}

\subsection{} Similar to generators the time is deterministic. the buffers will release the packets only at 0th, 1st, and second second. this is shown in Fig 9
\begin{figure}[bpht!]
\centering
	\includegraphics[scale=0.75]{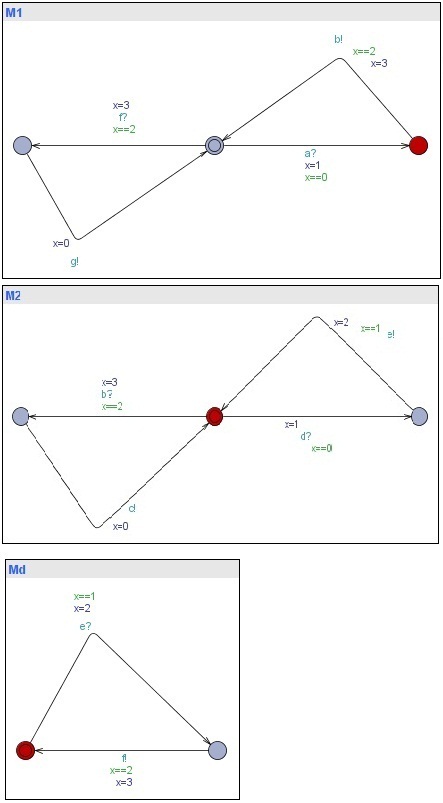}
		\caption{Buffers M1, M2 and Md in deterministic time}
	\label{Fig 9}
\end{figure}

\subsection{} Similar to generators and buffers the exits E1 and E2, release the packets at particular time as shown in the Fig 10 below.
\begin{figure}[bpht!]
\centering
	\includegraphics[scale=0.75]{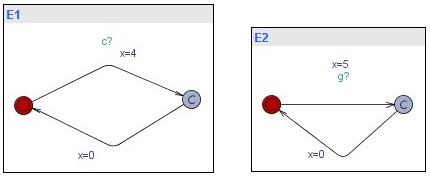}
		\caption{Exit E1 and E2 in deterministic time}
	\label{Fig10}
\end{figure}

\subsection{} The verification results for proposed model is shown in the Fig 11 below. The model is in Deadlock, because deterministic time approach violates the Two way concurrent buffer system timings where we require slight delay on or before a particular time, at various positions to make the system concurrent.
\begin{figure}[bpht!]
\centering
	\includegraphics[scale=0.70]{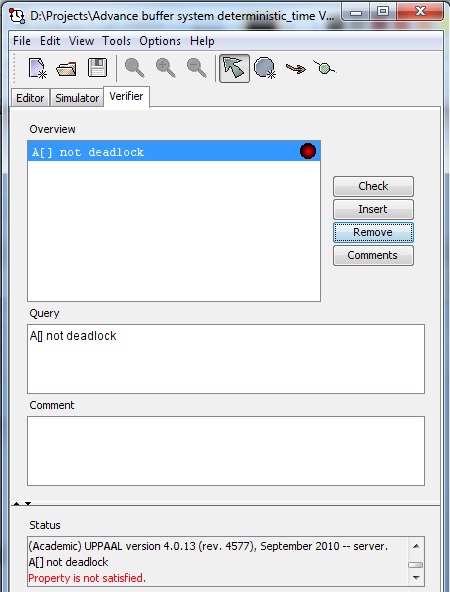}
		\caption{verification result for deterministic time}
	\label{Fig11}
\end{figure}

\section{Uppaal Model of Proposed Buffer System in Non-deterministic Time with no Deadlock}
\label{10}
Timing model on various states in the UPPAAL model for no deadlock is non deterministic time. in the figure below the model is incorporated with the time which is non deterministic. For the proposed two way buffer system, each component of the system is denoted by a separate automaton on the model.

\subsection{} G1 and G2 has been modeled as an automaton with two locations as shown in Fig 12. In G1 on transition, the packet is sent over the channel 'a' by the send operation a!, with a guard condition of $x \geq 0$ and invariant $x\leq 5$ when it moves to next location. So that next packet will not be produced until that time and once the packet is transferred through all the components, the guard condition is checked for $x\geq 5$. In G2 on transition, the packet is sent over the channel 'd' by the send operation d!, with a update of x=0, when it moves to next location and checks for invariant $x\leq 5$ at that location.  This is depicted below in Fig 12.

\begin{figure}[bpht!]
\centering
	\includegraphics[scale=0.90]{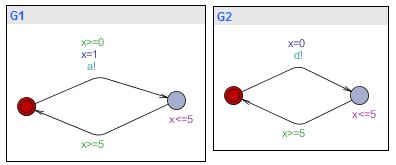}
		\caption{Generators G1 and G2  for non deterministic time}
	\label{Fig12}
\end{figure}

\subsection{} The model of buffer M1 consists of three locations. Firstly, when it receives a packet 	from G1 over the channel 'a' by the receive operation a?, with a guard condition $x\geq 0$ and invariant $x\leq 3$ it reaches to the next location. The packet received and is ready to be sent over the channel 'b' by the send operation b! to buffer M2. This is done with a guard condition $x>2$ and it updates the clock time to $x=3$. Similarly when it receives packet from M2 through the channel 'f' by receive operation f?. The packet is then sent to E2 through the channel 'g' through send operation g!. Fig 13 depicts the model of Buffer M1.

\begin{figure}[bpht!]
\centering
	\includegraphics[scale=1.0]{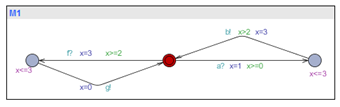}
		\caption{Buffer M1 for non deterministic time}
	\label{Fig13}
\end{figure}
	
\subsection{} Buffer M2 is also modeled on similar lines but with different invariant constraints which is shown in Fig 14.

\begin{figure}[bpht!]
\centering
	\includegraphics[scale=1.0]{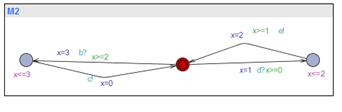}
		\caption{Buffer M2 for non deterministic time}
	\label{Fig14}
	\end{figure}
	
\subsection{} The dummy buffer $M_{d}$ , Clock variable, E1 and E2 are modeled as shown in Fig 15.

\begin{figure}[bpht!]
\centering
	\includegraphics[scale=1]{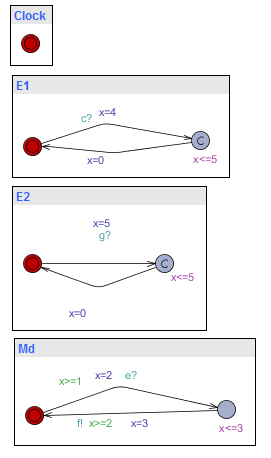}
		\caption{Clock, Exits E1 and E2, Dummy buffer Md for non deterministic time}
	\label{Fig15}
\end{figure}

\subsection{} The simulation results of the proposed two way concurrent system given by UPPAAL are shown in the Fig 16 where the packet goes in the same sequence as describes in section VII. In UPPAAL, the simulation has two ways to start, in first way G1 initiate the packet after that G2 which is shown in Fig 16 (a), in second way G2 initiate the packet after that G1, which is shown in Fig 16 (b).
\begin{figure}[bpht!]
\centering
	\includegraphics[scale=0.65]{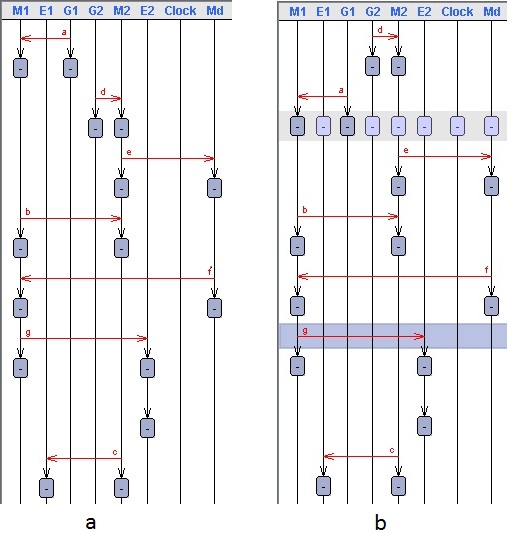}
		\caption{Simulation results for Two way concurrent buffer system for non deterministic time}
	\label{Fig16}
\end{figure}

We have modeled the proposed two way buffer system in UPPAAL. The main purpose of this work is to verify the model with respect to a requirement specification. Like the model, the requirement specification is expressed in a formally well-defined language \cite{MHMR04}. UPPAAL tool has been used to test three properties \textit{reachability}, \textit{safety (Deadlock)} and \textit{liveness}.

\begin{enumerate}[(a).]
	\item \textbf{Reachability:} The reachability property is described as 'Does there exist a path starting at the initial state, such that $\varphi$ is eventually satisfied along that path' \cite{BDL06}. This specification in UPPAAL is expressed as \textbf{\textit{E$<>$ M1.G2\_receive}}. This property checks for a path from M1 to G2 and the proposed model satisfies this property. The verification result of reachability property is shown in Fig 17.

\begin{figure}[bpht!]
\centering
	\includegraphics[scale=0.75]{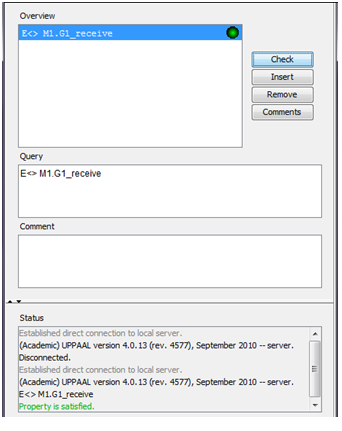}
	\caption{Reachability Property satisfied for non deterministic time}
	\label{Fig17}
\end{figure}
	
	\item \textbf{Safety:} Safety property is of the form 'For the path formula $\varphi$ covering all the states (locations) there is no deadlock in the system'. The safety specification in UPPAAL is expressed as \textbf{\textit{A[] no deadlock}}. This property checks for deadlock and the proposed model satisfy this property. The Fig 18 depicts the  satisfiability of the safety property. 
  
\begin{figure}[bpht!]
\centering
	\includegraphics[scale=0.75]{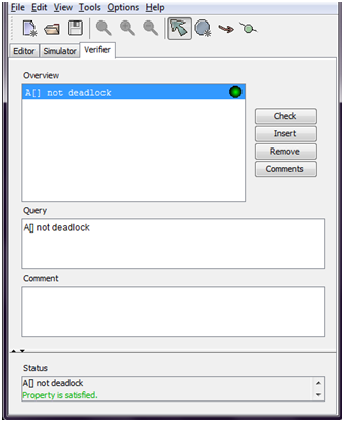}
		\caption{Safety Property satisfied for non deterministic time}
	\label{Fig18}
\end{figure}
	
	\item \textbf{Liveness:} In the model when a packet is sent it should eventually be received. This specification in UPPAAL is expressed as \textbf{\textit{G1.send $--$$>$ E1.receive}}. It checks whether E1 is eventually receiving the packet sent from G1 and in the result which is shown in Fig 19 this property is satisfied. 

\begin{figure}[bpht!]
\centering
	\includegraphics[height=6cm, width=8cm]{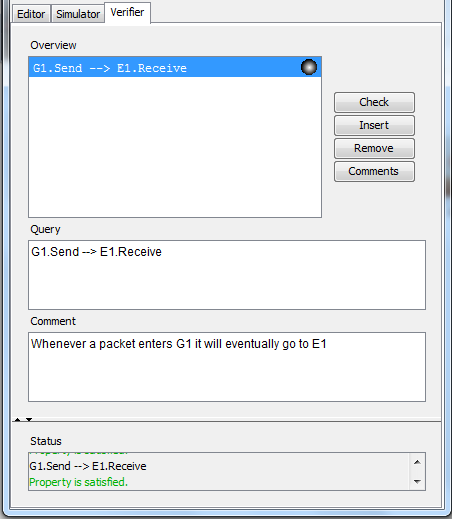}
		\caption{Liveness Property satisfied}
	\label{Fig19}
\end{figure}
\end{enumerate}

\section{Buffer Sizing}
\label{11}
Buffer system widely used in routers, Buffer sizing is very important aspect of the buffer system. As the proposed Two Way Buffer System may be over buffered in a high speed network system and be unusable and can lead to many complaints of congestion \cite{REF1}. But keeping aspects of buffer sizing standards can reduce the chances of congestion or any other flaws in high speed networks too.  The paragraph below is giving light on the buffer sizing constraints.

\par Internet routers are packet switches, and therefore buffer packets during times of congestion.  Arguably, router buffers are the single biggest contributor to uncertainty in the Inter-net. Buffers cause queuing delay and delay-variance; when they overflow they cause packet loss, and when they under-flow they can degrade throughput. Given the significance of their role, we might reasonably expect the dynamics and sizing of router buffers to be well understood, based on a well-grounded theory, and supported by extensive simulation and experimentation \cite{REF1}.

\par Router buffers are sized today based on a rule-of-thumb commonly attributed to as discussed in \cite{REF3} . Using experimental measurements of at most eight TCP flows on a 40 Mb/s link, they concluded that because of the dynamics of TCP's congestion control algorithms  a router needs an amount of buffering equal to the average round trip time of a flow that passes through the router, multiplied by the capacity of the router's net-work interfaces. 
\par This is the well-known \cite{REF1}
\begin{equation}
B = RTT * C_{rule}
\end{equation}

\par Network operators follow the rule-of-thumb and require that router manufacturers provide 250ms (or more) of buffering \cite{REF4}. The rule is found in architectural guidelines \cite{REF5}, too. Requiring such large buffers complicates router design, and is an impediment to building routers with larger capacity. For example, a 10Gb/s router line card needs approximately 50ms�10Gb/s= 2.5Gbits of buffers, and the amount of buffering grows linearly with the line-rate \cite{REF2}.

\par It is not well understood how much buffering is actually needed, or how buffer size affects net-work performance \cite{REF6}.

\par Overall  study of buffer sizing lights on the usability of small buffers in the high speed network.  significantly smaller buffers could be used in backbone routers  without a loss in network utilization \cite{REF1}.

\section{Scope for Future Enhancement}
\label{12}
The Two way buffer system can be used in the backbone of the network where two way transfer of the packet is required on a single line. The two way system will provide slight delay in the channel and support for concurrent packet transfer. The buffer system mainly used in routers. There are various kind of routers available in the market which fulfills different purposes. The Two way buffer system is most appropriate for core routers which are used in backbone of the network, subscriber edge routers this type of router belongs to an end user (enterprise) organization. The Two Way Buffer System is also appropriate for the system where there is heavy traffic of packets from one way and less traffic from other way. 
\par This research is beyond the scope of Ultra high speed network (GBPS) where optical links are used to transfer the data. This can be done by making "Two Way Concurrent Buffer System" more specific to this kind of networks, by keeping constraints of such system in mind.

\section{Conclusion}
\label{13}
The existing two way tiny buffer system when modeled using the UPPAAL tool it satisfy only the safety property only. The system is not time efficient because of unusual delays in transmission of packets of some buffer locations. In this paper we have tried to overcome the drawback of the existing tiny two way buffer system by incorporating a dummy buffer Md. The entire model has been modeled and verified in UPPAAL which incorporates the concepts of Timed Automata. The simulation result has satisfied all the three properties namely Safety, Liveness and Reachability, also it is fully concurrent.


\bibliographystyle{IEEEtran}
\balance
\bibliography{ref}
\end{document}